\documentclass[a4paper,10pt]{article}
\usepackage{enumerate}
\usepackage{color}
\usepackage[utf8]{inputenc} 
\usepackage[english]{babel}
\usepackage[T1]{fontenc}
\usepackage{graphicx}
\usepackage{amsfonts,amssymb,amsmath,latexsym,amsthm}
\usepackage{textcomp}
\usepackage[pdftex]{hyperref}
\usepackage{geometry}
\DeclareMathOperator{\sech}{sech}

\title{Solitary waves on flows with an exponentially sheared current and stagnation points}
\author{Marcelo V. Flamarion$^1$ and Roberto Ribeiro-Jr$^2$}
\date{}

\begin{document}
\maketitle
\begin{center}
{\footnotesize $^1$Unidade Acad{\^ e}mica do Cabo de Santo Agostinho\\
	UFRPE/Rural Federal University of Pernambuco \\
	BR 101 Sul, Cabo de Santo Agostinho-PE, Brazil,  54503-900 \\
marcelo.flamarion@ufrpe.br }

\vspace{0.3cm}
{\footnotesize $^2$Departament of Mathematics\\
   UFPR/Federal University of Paran{\' a} \\
   Centro Polit\'ecnico, Jardim das Am\'ericas, Curitiba-PR, Brazil, 81531-980\\
   robertoribeiro@ufpr.br
}

%{\footnotesize ORCID Number: 0000-0002-0877-4831}

\end{center}

\begin{abstract}
\noindent While  several   articles have been written on water waves on flows with constant vorticity, little is known about  the extent to which a nonconstant vorticity affects the flow structure, such as the appearance of  stagnation points. In order to shed light on this topic,  we investigate in  detail the flow beneath   solitary waves propagating on an exponentially decaying sheared current. Our focus is to analyse numerically the emergence of stagnation points. For this purpose, we approximate the velocity field  within the fluid bulk through the classical Korteweg-de Vries asymptotic expansion and use the Matlab language to evaluate the resulting streamfunction. Our findings suggest  that the flow beneath the waves can have zero, one or two stagnation points in the fluid body.  We also study the bifurcation  between these flows.  Our simulations indicate that the stagnation points emerge from a streamline with a sharp corner.
\end{abstract}


%\maketitle

\section{Introduction}
\label{sec_intro}
In the last three decades a great attention has been given to the study of waves propagating on linearly sheared currents (constant vorticity). Many advances have been made in a better understand of the effects of this kind of   current on the free surface and in the bulk of the fluid. %It is noteworthy to mention the contribution of the  research groups leadered by Constantin and Vanden-Broeck in this field.  

It is well-known that the flow beneath a rotational water wave  with constant vorticity is characterized mainly by: i) the appearance of stagtnation points -- fluid particles with zero velocity in the wave's moving frame; ii) critical layers --  a horizontal layer with closed streamlines separating the fluid into two disjoint regions; iii) the arising of pressure anomalies such as  the occurrence of   maxima and minima of pressure in locations other than below the crest and trough respectively. These results have been showed numerically \cite{Choi03,RobertoPaulAndre,VasanOliveras,TelesPeregrine}, asymptotically \cite{AliKalish,Johnson86}    and also proved rigorously \cite{EhrnstromVillari,Wahlen,Kozlov}. The constant vorticity also affects the shape of the free surface, for instance  its profile can become rounder and possibly form an exotic overhanging wave  \cite{Vanden-Broeck94,Vanden-Broeck96,DyachenkoHur19a,DyachenkoHur19b}.

Although many realistic situations can be represented by constant vorticity (e.g. when waves are long compared with the water depth or when waves are short compared with lengthscale of the vorticity distribution \cite{TelesPeregrine}) there are  cases that cannot. For instance, when the current is generated by wind stress at the surface   a highly sheared current near the water surface  with non-uniform vorticity results \cite{Swan}.

The study of waves with arbitrary vorticity distribution started in 1934 with Dubreil-Jacotin \cite{DJ}.  She introduced a change of variables that   transforms the domain of the fluid into a rectangle, then by using power series she  computed  small-amplitude  waves  with general vorticity. Since then, a number of authors have studied  this problem by considering  different current profiles (vorticity distribution) and approaches.

Starting from the results on infinitesimal rotational waves, we mention the works of Peregrine \cite{Peregrine76} and Dalrymple\cite{Dalrymple}. In these works the authors   present a detailed discussion on   simplified models for currents used in several  studies (currents that  are modelled as a bilinear,  a sine, a cosine or  a one-seventh power law as function of the depth). % Besides, \cite{Dalrymple} present a numerical study for hfjdkfshkfsh.

In an attempt of moving on the direction of nonlinear waves,  Freeman and Johnson \cite{FreemanJohnson} deduced a Korteweg-de Vries equation  for waves on the surface of a general sheared current.      A rigorous theoretical result on the existence of large-amplitude waves with general
vorticity distributions   was given only 70 years after Dubreil-Jacotin \cite{DJ} findings in the seminal work of  Constantin and Strauss \cite{ConstantinStrauss04}.  Their results relies on the hypothesis of periodic waves over a  finite depth channel and vorticity-stream-function in $C^{1,\alpha}$. This work opened the way to several theoretical and numerical fruitful researches in the coming years. Among them, we mention the findings of Hur \cite{Hur08} and  Groves and Wahl\'en \cite{GrovesWahlen} who  proved  the existence of solitary waves of small amplitude for an arbitrary vorticity.  Besides, the contributions of Hur \cite{Hur11} who proved the existence of large-amplitude  periodic  waves on deep water regime.  

Due to technical complications, the numerical investigations for large-amplitude waves with nonconstant vorticity  have been conducted under the premise  that there is no stagnation point in the bulk of the fluid. 
Among the works on this topic, we highlight the findings of   Ko and Strauss \cite{KoStrauss08a,KoStrauss08b}. These authors computed   periodic water waves for:  a shear vorticity layer of some uniform thickness,  a continuous vorticity  and for  the case in which the vorticity is represented by a step function.  Their studies focused on certain characteristic of the flow such as the  probable location of the first stagnation point and the shape of the nearly stagnant waves. Later,  Amann and Kalimeris \cite{AmannKalimeris18} improved the results of Ko and Strauss \cite{KoStrauss08a,KoStrauss08b}  in the sense that they  computed waves closer to stagnation. They  also constructed waves for flows with  vorticity that  varies linearly and quadratically with respect the stream function. More recently, Chen et al. \cite{Chen et al} presented an complete description of the flow beneath periodic water waves propagating on rotational flows with piece-wise constant vorticity. They  took into account the effects of the  vorticity on the free surface, on the streamlines, on the pressure distribution in the fluid and also showed particle trajectories.

A full description of the flow pattern beneath a large-amplitude wave with nonconstant vorticity in the presence of internal stagnation points is still an opened problem.  The state of the art of this topic is restrict to linear waves  and for the single case in which the vorticity depends linearly upon the stream function.  Regarding this subject, we refer to the study of  Ehrnsrtr{\"o}m et al. \cite{EhrnstromEscherVillari} in which the authors   give a qualitative description of the cat's eye structure. They show that  arbitrarily many critical layers with cat's vortices are possible  beneath linear waves with  an affine vorticity. This is also proved by Ehrnsrtr\"om et al. \cite{EhrnstromEscherWahlen} whose work is later revisited  by Aasen and Varholm \cite{AasenVarholm} through a slightly different approach.   

In this work we study numerically the flow structure beneath  solitary waves propagating on an exponentially sheared current. Our goal is to capture with details the appearance of  stagnation points.  For this purpose, we consider the classical Korteweg-de Vries model to obtain an approximation of the velocity field in the fluid body.   We find parameter regimes in which the flow has zero, one or two stagnation points. Differently from the constant vorticity case, in which the  stagnation points gives rise to the  Kelvin's cat's eye' flow pattern,  now they form a  flow  characterized  by  a  ``loop-pattern'', namely, there is a closed flow region with its corresponding separatrix is in the shape of a loop.  By varying the parameters of the flow, we show  that the two stagnation points (a centre and a saddle) that define the loop-pattern get closer, until they  collide, then  vanish, resulting in a saddle-centre bifurcation.  Besides, we notice that  a  streamline with a sharp corner results from the collision of the  two stagnation points.

This paper is organized as follows. In section  \ref{sec_eq_evol} we revisit the classical  KdV equation for waves on  a general sheared current  and present the  approximation of the velocity field.  At the  beginning of the section    \ref{trajetorias}, we  discuss   the ODE system for the particle  trajectories. Subsequently, the numerical approach and the results  are shown.  Concluding remarks are given in section \ref{conclusao}.

%###################################
\section{Governing equations}
\label{sec_eq_evol}
Consider a two-dimensional incompressible flow of an inviscid fluid of constant density $\rho$ in a finite channel with depth $h_0$. Denote by $(u(x,y,t),v(x,y,t))$ the velocity field in the bulk of fluid and $\zeta(x,t)$ the free surface profile. Choosing the set of dimensionless variables, where $\lambda$ is a typical wavelength, $a$  is a typical wave amplitude, $h_0$ is  a typical  depth and ${\lambda}/{\sqrt{gh_{0}}}$ is a typical time-scale we have the dimensionless governing Euler equations
\begin{align} \label{K2}
	\begin{split}
		& u_{t} + uu_{x} + vu_{y} = - p_{x}, \;\  \mbox{for} \;\  0 < y < 1 +\epsilon\zeta(x,t), \\
		&  \mu^{2}\{v_{t} + \epsilon(uv_{x} + vv_{y})\} = - p_{y}, \;\  \mbox{for} \;\  0 < y < 1 +\epsilon\zeta(x,t), \\
		&  u_{x} + v_{y} = 0, \;\  \mbox{for} \;\  h(x) < y < 1 +\epsilon\zeta(x,t), \\
		& p=  \epsilon\zeta, \;\ \mbox{at} \;\ y =1 +\epsilon\zeta(x,t),\\
		& v = \epsilon(\zeta_{t} + u\zeta_{x}),  \;\ \mbox{at} \;\ y = 1 +\epsilon\zeta(x,t), \\
		& v =0, \;\ \mbox{at} \;\ y = 0, \\
	\end{split}
\end{align}
where $\mu = {h_{0}}/{\lambda}$ is the dispersive parameter and  $\epsilon = {a}/{h_{0}}$ is the nonlinearity parameter.

We assume that the flow is in the presence of a depth-dependent imposed current $(U(y),0)$, which dominates the the velocity field. We are interested in investigating particle trajectories in the bulk of the fluid beneath a solitary wave. To this end, we consider the asymptotical approximation for the velocity field as well as for the free surface reported by  Johnson \cite{JohnsonBook}. In order to ease the read of the article, we summarise bellow the main steps necessary to obtain the reduced system of governing equations. 

We consider the scaling
\begin{align}\label{s1} 
	\begin{split}
		u\rightarrow U(y) + \epsilon u, \;\ v\rightarrow \epsilon v, \;\ p \rightarrow\epsilon p, 
	\end{split}
\end{align}
and  the standard long-wave approximation $\mu^{2} = \epsilon$. The condition (\ref{s1})  yields 
\begin{align*}% \label{K3}
	\begin{split}
		& u_{t} + U(y)u_{x} + U'(y)v + \epsilon(uu_{x} + vu_{y}) = - p_{x}, \;\ \mbox{for} \;\  0 < y < 1 +\epsilon\zeta(x,t), \\
		&  \epsilon\{v_{t} + U(y)v_{x} + \epsilon(uv_{x} + vv_{y})\} = - p_{y}, \;\  \mbox{for} \;\  0 < y < 1 +\epsilon\zeta(x,t),  \\
		&  u_{x} + v_{y} = 0,  \;\  \mbox{for} \;\  0 < y < 1 +\epsilon\zeta(x,t), \\
		& p=\zeta, \;\ \mbox{at} \;\ y = 1 +\epsilon\zeta,\\
		& v = \zeta_{t} + U(y)\zeta_{x} + \epsilon u\zeta_x, \;\ \mbox{at} \;\ y = 1 +\epsilon\zeta, \\
		& v =0, \;\ \mbox{at} \;\ y = 0. \\
	\end{split}
\end{align*}
Introducing the  variables $\xi = x - ct$ and $\tau = \epsilon t$, where $c$ is to be determined {\it a posteriori}, we obtain
\begin{align} \label{K4}
	\begin{split}
		& (U(y)-c)u_{\xi} +U'(y)v + \epsilon(u_{\tau} + uu_{\xi} + vu_{y})= - p_{\xi}, \;\  \mbox{for} \;\  0 < y < 1 +\epsilon\zeta,  \\
		& \epsilon\{(U(y)-c)v_{\xi} + \epsilon(v_{\tau} + uv_{\xi} + vv_{y})\}= -p_{y}, \;\  \mbox{for} \;\  0 < y < 1 +\epsilon\zeta,  \\
		& u_{\xi} + v_{y} = 0, \;\  \mbox{for} \;\  0 < y < 1 +\epsilon\zeta,  \\
		& p=\zeta , \;\ \mbox{at} \;\ y = 1 +\epsilon\zeta,\\
		& v = (U(y)-c)\zeta_{\xi} + \epsilon(\zeta_{\tau}+u\zeta_{\xi}),  \;\ \mbox{at} \;\ y = 1 +\epsilon\zeta, \\
		& v =0 \;\ \mbox{at} \;\ y = 0. \\
	\end{split}
\end{align}

Seeking for asymptotic solutions of (\ref{K4}), we define the power series expansion
\begin{equation}\label{K5}
	q(\xi,y,\tau;\epsilon) = \sum_{n=0}^{\infty}\epsilon^{n}q_{n}(\xi,y,\tau), \;\ \zeta(\xi,\tau) = \sum_{n=0}^{\infty}\epsilon^{n}\zeta_{n}(\xi,\tau), 
\end{equation}
where $q=u, v$ or $p$. Substituting  (\ref{K5}) in (\ref{K4}) and procceding  as in \cite{JohnsonBook} we obtain as first approximation for the velocity field in bulk of the fluid
\begin{equation}\label{K7}
	v_{0} = \bigg\{(U(y)-c)\int_{0}^{y}\frac{dz}{(U(y)-c)^2}\bigg\}\zeta_{0\xi},
\end{equation}
\begin{equation}\label{K8}
	u_{0} = -\bigg\{ \frac{1}{U(y)-c}+U'(y)\int_{0}^{y}\frac{dz}{(U(y)-c)^2}\bigg\}\zeta_{0},
\end{equation}
where $c$ satisfies the compatibility condition known as Burns condition:
\begin{equation}\label{Burns}
	\int_{0}^{1}\frac{1}{(U(y)-c)^2} dy= 1,
\end{equation}
and the free surface $\zeta_0$ is solution of the sheared Korteweg-de Vries equation
\begin{equation}\label{KdV}
	-2I_{31}\zeta_{0\tau}+3I_{41}\zeta_{0}\zeta_{0\xi}+J_{1}\zeta_{0\xi\xi\xi} = 0,
\end{equation}
whose coefficients $I_{n1}=I_{n}{(1)}$, $n=1,2,3,4,$ are given by
\begin{equation}\label{In}
	I_{n}(y )\equiv\displaystyle\int_{0}^{y}\frac{dy}{(U(y)-c)^{n}},
\end{equation}
and
\begin{equation}\label{J1}
	J_{1} = \displaystyle\int_{0}^{1}\int_{y}^{1}\int_{0}^{y_1}\frac{(U(y_{1})-c)^{2}}{(U(y)-c)^{2}(U(y_{2})-c)^{2}}dy_{2}dy_{1}dy. 
\end{equation}

Solitary wave solutions of (\ref{KdV}) are described by the formula
\begin{equation}\label{solitary}
	\zeta_{0}(\xi,\tau)=A\sech^{2}(k(\xi-C\tau)), \mbox{ where } k=\sqrt{\frac{AI_{41}}{4J_1}} \mbox{ and } C=-\frac{2J_1k^2}{I_{31}}.
\end{equation}
In the Euler coordinates, the solitary wave solution is written as
\begin{equation}\label{solitaryEuler}
	\zeta_{0}(x,t)=\epsilon A\sech^{2}\Big(k(x-(c+\epsilon C)t)\Big), \mbox{ where } k=\sqrt{\frac{AI_{41}}{4J_1}} \mbox{ and } C=-\frac{2J_1k^2}{I_{31}}.
\end{equation}

%%%%%%%%%%%%%%%%%
\section{Computing particle trajectories} \label{trajetorias}

Particle trajectories beneath the solitary wave (\ref{solitaryEuler}) can be approximated by the solution of the system of ordinary differential equations (ODEs)
\begin{align} \label{DSf}
	\begin{split}
		& \frac{dx}{dt} = U(y)  + \epsilon u_{0}(x,y,t), \\
		& \frac{dy}{dt} =\epsilon v_{0}(x,y,t). \\
	\end{split}
\end{align}

It is convenient to solve (\ref{DSf}) in the wave frame $X= x-(c+\epsilon C)t$ and $Y=y$. In this framework, particle trajectories are solutions of the autonomous ODE system
\begin{align} \label{DS}
	\begin{split}
		& \frac{dX}{dt} = U(Y)- (c+\epsilon C)  + \epsilon u_{0}(X,Y), \\
		& \frac{dY}{dt} =\epsilon v_{0}(X,Y). \\
	\end{split}
\end{align}

It is easy to see that streamlines are solutions of (\ref{DS}). Thus,  particle trajectories are the level curves of the stream function $\Psi(X,Y)$, which is given by
\begin{equation} \label{streamfunction}
	\Psi(X,Y) = \int_{0}^{Y}\left(U(s)+\epsilon u_{0}(X,s) \right) \,ds -(c+\epsilon C)Y.
\end{equation}
In order to compute the streamlines (\ref{streamfunction}) we need to solve the integrals (\ref{Burns}), (\ref{In}) and (\ref{J1}). These integrals are obtained through the function {\it integral} that is implemented in {MATLAB}. Following we present the results for a linear and an exponentially vertically sheared current.

\subsection{Linear current}\label{linear_corrente}

Guan \cite{China} studied particle trajectories beneath solitary waves in the presence of a linear sheared current through the classical asymptotic long-wave limit. He showed that the orbits obtained from the asymptotic approximation agree well with the ones produced by full Euler equations when the wave amplitude is small. Based on his results in all simulations presented in this article we fix $\epsilon =0.1$.

\begin{figure}[h!]
	\centering
	\includegraphics[scale=1]{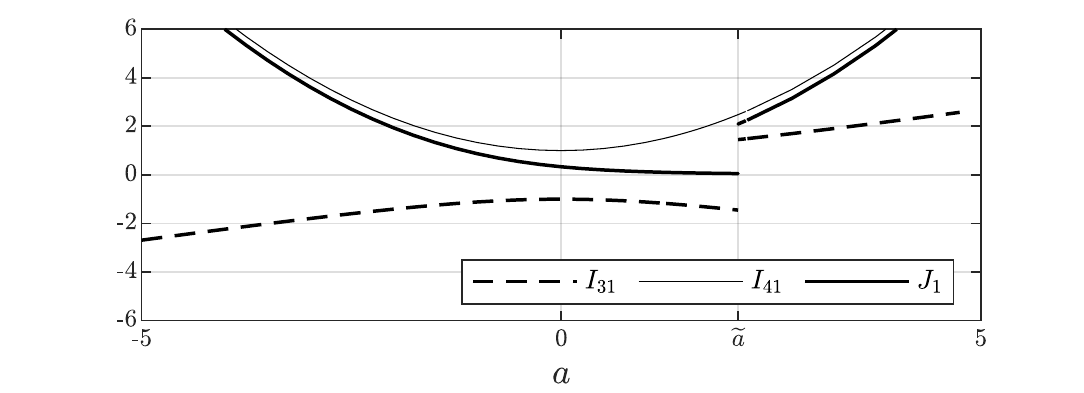} 
	\caption{Coefficients of KdV in terms of $a$ for the  linear sheared  current ($\widetilde{a} \approx 2.1092$).}
	\label{Coeficientes_KdV_linear}
\end{figure}

\begin{figure}[h!]
	\centering
	\includegraphics[scale=1]{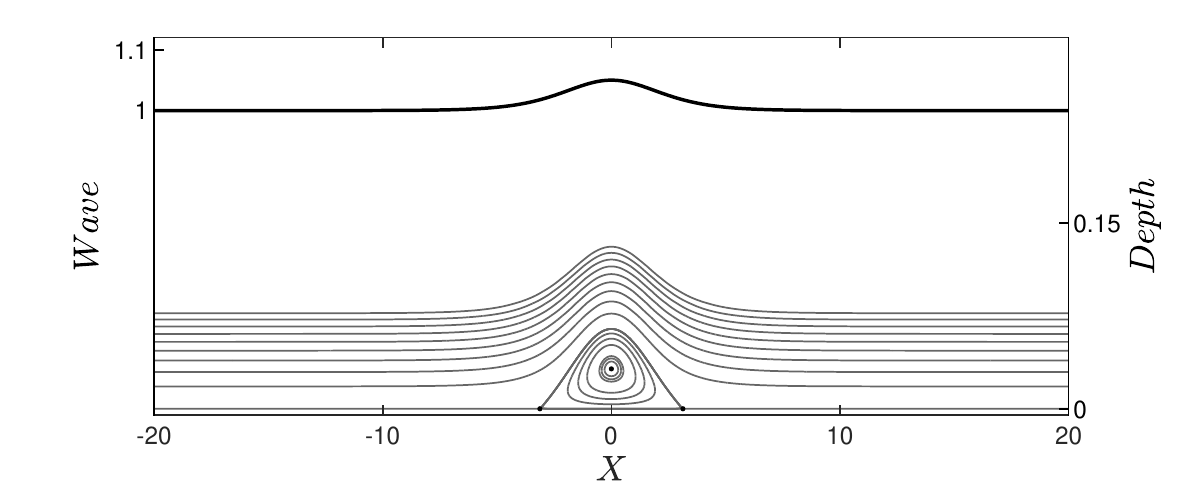} 
	\caption{Typical phase portrait of  (\ref{DS}) for a linear sheared current. }
	\label{Comp1}
\end{figure}

In order to verify our mathematical formulation for the particle trajectories, we start considering the  linear sheared current $U(y)=ay$ and analyse how the parameters of the KdV (\ref{KdV}) varies according to the value of  $a$. This is shown in figure \ref{Coeficientes_KdV_linear}. We see that for positive values of   $a$  the coefficients $J_{1}$ (that controls the  dispersion) and $I_{41}$ (that controls the nonlinearity effects)  reproduce different behaviours. While $J_1$ is decreasing  in the interval $(0,\widetilde{a})$  and  approaches to zero as $a \to \widetilde{a}$, $I_{41}$ is increasing in the same interval. Furthermore, we notice that  $J_1$ and $I_{41}$   have a  discontinuity at $a = \widetilde{a}$. This shows us  that  the KdV model is not appropriate to study flows  with $a\gg0$. With this in mind, we reproduce a typical phase portrait for $a<0$. Figure \ref{Comp1} shows the phase portrait of the ODE (\ref{DS}) for $U(y)=-30y$. This case is featured by the existence of a critical layer close to the bottom with three stagnation points, two saddles and a centre, with the centre beneath the crest of the wave --  and this ties in with the results in the literature  \cite{China, TelesPeregrine, RobertoPaulAndre}.

Our approach allow us to go beyond constant vorticity flows and study the existence of stagnation points in the presence of more complex types of currents such as exponentially decaying ones.

\subsection{Exponentially decaying current}\label{correnteza_exponencial}

\begin{figure}[h!]
	\centering
	\includegraphics[scale=1]{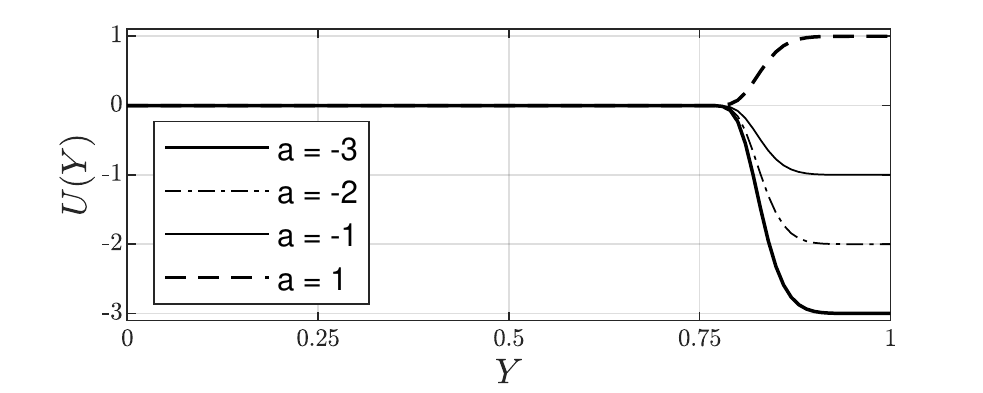} 
	\caption{Current profile $U(y)=a\exp\Big(-\sigma(y-1)^{8}\Big)$ for different values of $a$ ($\sigma = 10^6$).}
	\label{Profile_U}
\end{figure}

An interesting physical problem in water waves over a vertical variable current is surface shear waves  \cite{Peregrine74}. Surface shear wave flows are idealized
assuming a thin fast-moving sheet of fluid at the surface and still water beneath it. In order to investigate particle trajectories under these conditions, we choose as the current profile the function 
\begin{equation}\label{exp}
	U(y)=a\exp\Big(-\sigma(y-1)^{8}\Big), 
\end{equation}
where $a\in\mathbb{R}$ and $\sigma>0$ is chosen so that the horizontal velocity decays rapidly with the depth.  In figure \ref{Profile_U} we depict  the profile of the current (\ref{exp})  for some values of $a$. In the simulations presented here, we consider $\sigma = 10^6$. This choice of parameters represent a current acting roughly in $20\%$ of the water column.

\begin{figure}[h!]
	\centering
	\includegraphics[scale=1]{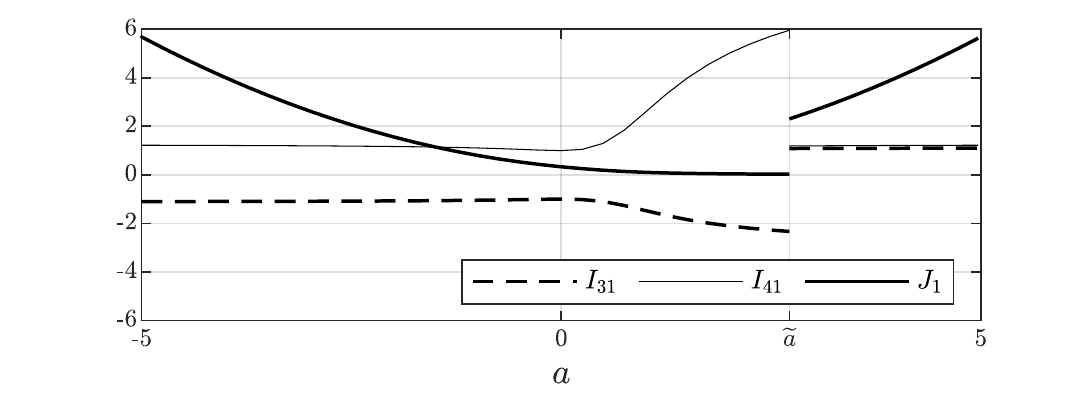} 
	\caption{Coefficients of KdV in terms of $a$ for the exponentially decay current ($\widetilde{a} \approx 2.7207$).}
	\label{Coeficentes_KdV_exp}
\end{figure}

Figure \ref{Coeficentes_KdV_exp} displays  the graph of parameters of the KdV (\ref{KdV}) as a function of $a$ for the exponentially decaying current (\ref{exp}). As in the linear case, we notice that the KdV model is not appropriate for values of $a\gg0$. 

For flows with  constant vorticity,  it is known that for $a\ll0$   there are three stagnation points: a centre  below the crest (inside the flow) and two saddles on the bottom, resulting in a cat's eye structure below the crest and attached to the bottom (see figure \ref{Comp1} or the figures in \cite{China}).  As $a$ increases the  three stagnation points coalesce at   $x=0$ and disappear in a saddle-centre bifurcation \cite{RobertoPaulAndre}.

\begin{figure}[h!]
	\centering
	\includegraphics[scale=1]{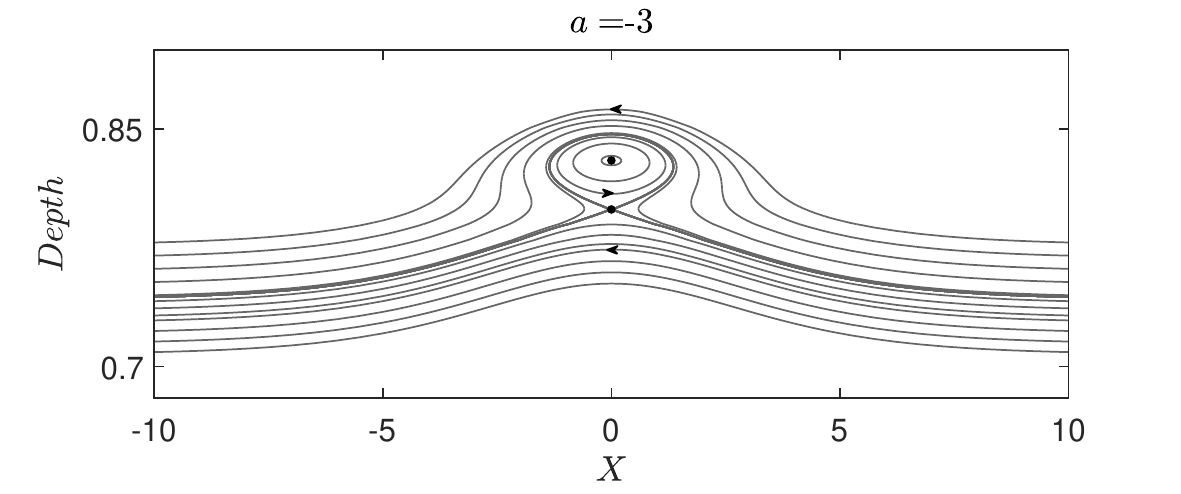} 
	\includegraphics[scale=1]{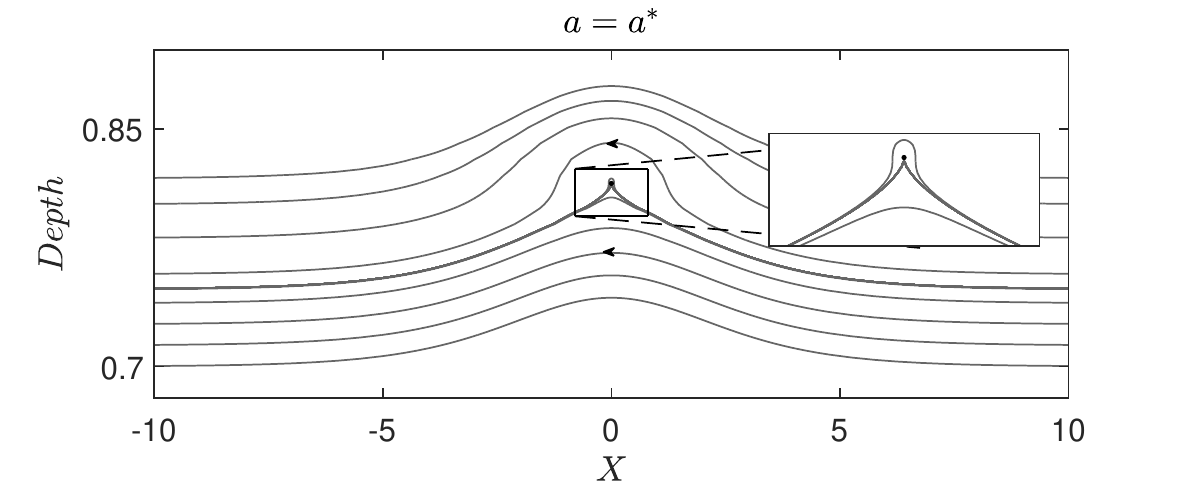} 
	\includegraphics[scale=1]{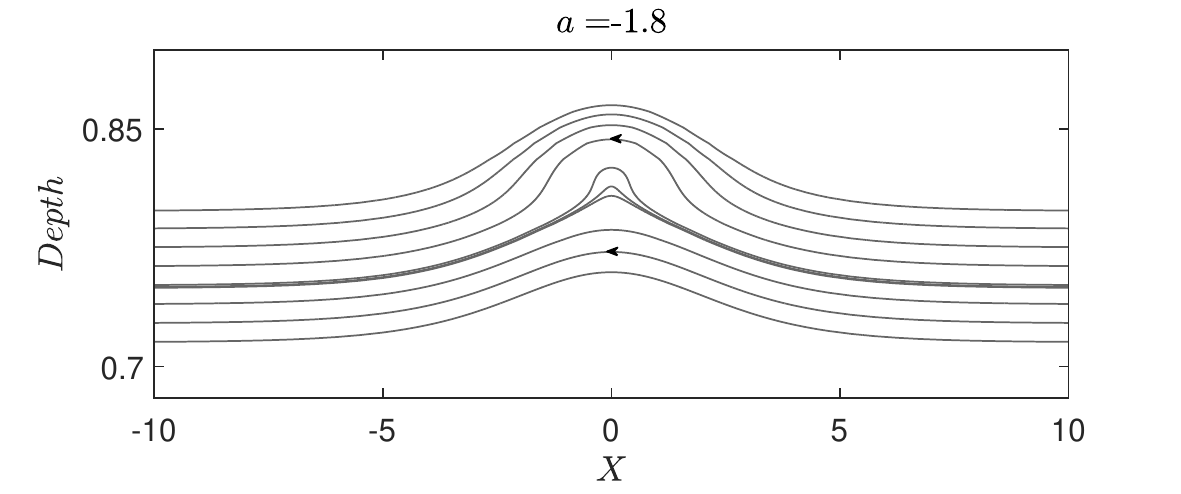} 
	\caption{Phase portraits for flows associated to the  exponentially decaying current (\ref{exp}) for different values of  $a$ ( $a^*\approx -1.897$).}
	\label{PhasePortraitExp}
\end{figure}

\begin{figure}[h!]
	\centering
	\includegraphics[scale=1]{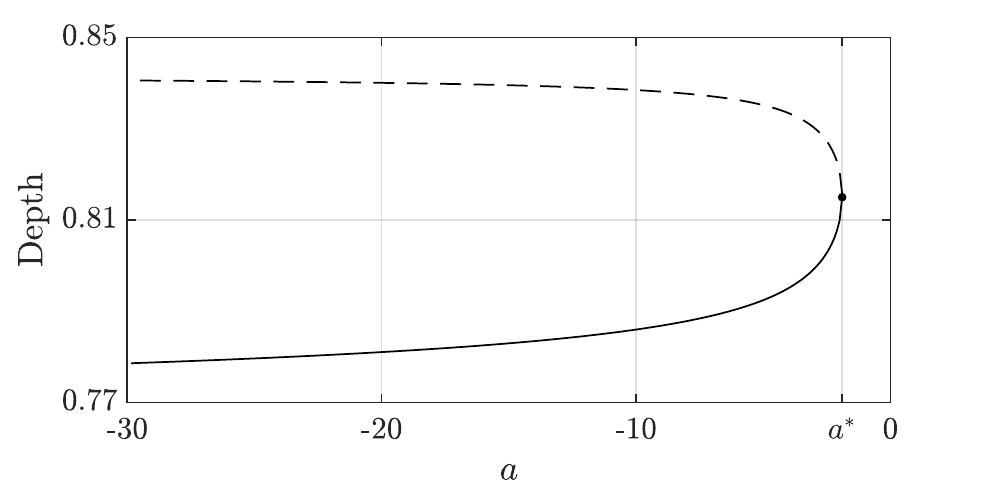} 
	\caption{Depth of the  centre point (dashed line) and the saddle point (solide line)  for different values of $a$ ($a^* \approx -1.897$) .}
	\label{Bifurcation}
\end{figure}

Differently from the constant vorticity case, for values of  $a\ll0$ the phase portrait of the flow associated  with the exponentially decaying current (\ref{exp}) has only two stagnation points:  one centre and one saddle -- both of them bellow the crest of the wave forming a recirculation zone,  see  figure \ref{PhasePortraitExp} (top).    As $a$ increases, the saddle moves up and the centre moves down  shrinking the recirculation zone, then these points coalesce  forming a single stagnation point at $a^* \approx -1.897$.  The phase portrait for $a=a^{*}$ indicates the existence of a  streamline with a sharp corner at $x=0$ where its located a stagnation point. This happens when the two stagnation points collide. The  saddle-centre bifurcation  occurs from this streamline. 
%It is worth noting that similar phase portraits are obtained even when the current acts in less than $1\%$ of the water column.  
Besides, it calls our attention the similarity between the phaseportrait for $a = -3$ and  the one illustrated by Escher et al. \cite{EscherMatiocMatioc}(Figure 3) in the  study of stratified water waves with density varying linearly along the streamlines. These pictures indicate that  somehow there might exist a certain correspondence between stratification and vorticity as a mechanism of formation of critical layers.
Figure \ref{Bifurcation} depict the saddle-centre bifurcation in a different perspective. For $a>a^*$ there are no stagnation points.  

\section{Conclusions}\label{conclusao}
In this work, the focus has been on describing the main features of the flow beneath a solitary wave on an exponentially sheared current. The velocity field within the fluid bulk was approximated by the classical weakly nonlinear and weakly dispersive regime. This allowed to compute the location of  stagnation points and the streamlines. Of particular interest we reported the following results. Flows may have a single stagnation point in the fluid body located at the top of a streamline with a sharp corner. By varying the intensity of the current we obtained a flow with two stagnation points (a centre and a saddle) and a recirculation zone closed by a separatrix  loop shaped.
The visualization of such features can hopefully provide valuable insights for both  theoretical and numerical  works on water waves on flows with variable vorticity. Besides, although the results presented here are just for an exponential sheared current, the numerical procedure can be easily  adjusted for other profiles of sheared currents.

\end{document}